\documentstyle[aps,epsf]{revtex}

\begin{document}

\draft

\title{Test of renormalization predictions for universal finite-size scaling
functions}
\author{Erik Luijten\thanks{Electronic address:
        erik.luijten@uni-mainz.de}}
\address{Department of Physics, Delft University of Technology, P.O. Box
         5046, 2600 GA Delft, The Netherlands}
\address{Max-Planck-Institut f\"ur Polymerforschung, Postfach 3148,
         D-55021 Mainz, Germany}
\address{Institut f\"ur Physik, WA 331, Johannes Gutenberg-Universit\"at,
         D-55099 Mainz, Germany\thanks{Present address.}}


\maketitle

\begin{abstract}
  We calculate universal finite-size scaling functions for systems with an
  $n$-component order parameter and algebraically decaying interactions. Just
  as previously has been found for short-range interactions, this leads to a
  singular $\varepsilon$-expansion, where $\varepsilon$ is the distance to the
  upper critical dimension. Subsequently, we check the results by numerical
  simulations of spin models in the same universality class. Our systems offer
  the essential advantage that $\varepsilon$ can be varied continuously,
  allowing an accurate examination of the region where $\varepsilon$ is small.
  The numerical calculations turn out to be in striking disagreement with the
  predicted singularity.
\end{abstract}

\pacs{05.70.Jk, 64.60.Ak, 64.60.Fr}

In order to analyze numerical results obtained by Monte Carlo or
transfer-matrix studies of phase transitions and critical phenomena, {\em
finite-size scaling\/}~\cite{barber} is a very widely used technique.  This
hypothesis, which was for the first time properly formulated by
Fisher~\cite{fisher-fss}, allows the extrapolation of the properties of finite
systems, which do not exhibit a phase transition, to the thermodynamic limit.
In 1982, Br\'ezin~\cite{brezin} achieved a breakthrough by showing that the
finite-size scaling laws can actually be derived from
renormalization-group~(RG) theory, provided that the RG equations are not
singular at the fixed point. This implies a breakdown of finite-size scaling
for dimensionalities $d \geq 4$ and consequently an expansion of the
finite-size scaling functions in powers of $\varepsilon=4-d$ is {\em
singular\/} at $\varepsilon=0$.  This rather surprising result was confirmed by
explicit calculations for the $n$-vector model in the large-$n$ limit.  In
addition, it follows from Ref.~\cite{brezin} that the finite-size scaling
relation for the free energy is a universal function depending only on two
nonuniversal metric factors, without an additional nonuniversal prefactor. This
result was derived from different arguments by Privman and
Fisher~\cite{privman84} and subsequently confirmed analytically for the
spherical model~\cite{sp85}.  Pioneering work~\cite{bzj,rudnick} then showed
that a field-theoretic calculation of finite-size scaling functions is actually
possible.  Specifically, Br\'ezin and Zinn-Justin~\cite{bzj} developed a
systematic $\varepsilon$-expansion for these functions.  Unlike the standard
expansion in powers of $\varepsilon$ for critical exponents and scaling
functions of bulk properties, one finds, for a fully finite geometry, an
expansion in powers of $\sqrt{\varepsilon}$. More recently, Esser {\em et
al.}~\cite{esser} introduced a promising perturbation approach at fixed~$d$
which is applicable below the critical temperature as well. However, here we
focus on the expansion in $\varepsilon$ and in particular on the singular
nature of this expansion.

\begin{figure}
\leavevmode
\centering
\epsfxsize 10cm
\epsfbox{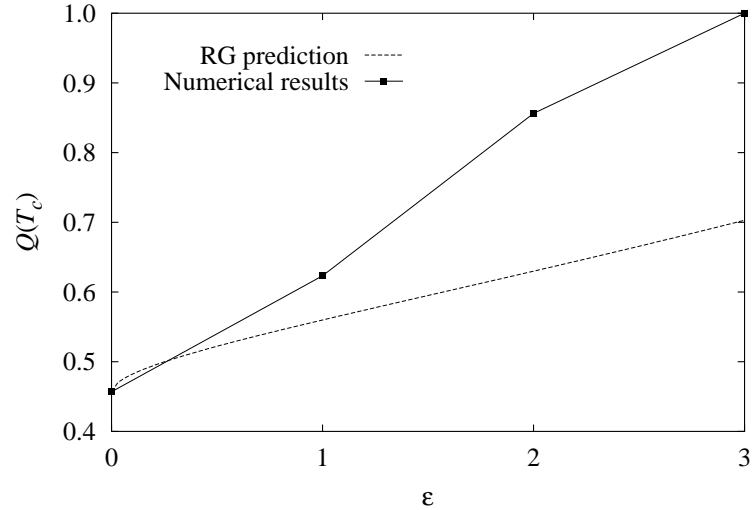}
\caption{The amplitude ratio $Q(T_c)$ for systems with short-range
interactions. The dashed line shows the $\protect\sqrt{\varepsilon}$-expansion
of Ref.~\protect\cite{bzj}.}
\label{fig:q-sr}
\end{figure}

The systems under consideration have an $n$-component order parameter~$\phi$
with ${\rm O}(n)$ symmetry and periodic boundary conditions.  A quantity of
central interest is the amplitude ratio $Q=\lim_{L \to \infty} \langle \phi^2_L
\rangle^2 / \langle \phi^4_L \rangle$, which is directly related to the
cumulant introduced by Binder~\cite{binder-cum}. At the critical temperature
$T_c$ it takes a universal, although geometry-dependent, value (cf.\ 
Ref.~\cite{privman84}). In Ref.~\cite{bzj}, an expansion for~$Q(T_c)$ was
obtained in powers of $\sqrt{\varepsilon}$, up to ${\cal O}(\varepsilon)$,
which is shown in Fig.~\ref{fig:q-sr} for $n=1$, along with numerical results
for integer~$d$.  Given the low order of the expansion and the fact that it can
only be checked for integer values of $\varepsilon$, hardly any conclusions can
be drawn from a comparison to numerical results and any confirmation of the
singular nature of the $\varepsilon$-expansion will have to wait until the RG
calculation has been carried to substantially higher order. Thus, we propose a
different route: namely, we replace the short-range (SR) forces by long-range
attractive interactions decaying as a power law, $J(r) \propto
r^{-(d+\sigma)}$, where $0<\sigma<2$. It has been shown in
Refs.~\cite{fmn,suzuki72} that these systems have an upper critical dimension
$d_c=2\sigma$ and that for $d<d_c$ the critical exponents can be calculated in
terms of an expansion in $\varepsilon' = 2\sigma-d$, very similar to the
original $\varepsilon$-expansion, which is recovered for $\sigma \to 2$.  Since
the upper critical dimension now is a continuous parameter, we have the
opportunity to verify $\varepsilon'$-expansion results for arbitrarily small
$\varepsilon'$.  So, even if actual physical realizations of this system may be
scarce, it constitutes a very valuable mathematical model. Interestingly, the
finite-size scaling functions for the spherical model with power-law
interactions have been calculated by several authors~\cite{lr-spherical} for
$d/2 < \sigma < d$, but the nature of a possible singularity in the limit
$\sigma \to d/2$ appears not to have been examined.  The first part of this
paper is therefore devoted to a generalization of the treatment of
Ref.~\cite{bzj} to systems with algebraically decaying interactions. For
notational convenience we redefine $\varepsilon=\varepsilon'=2\sigma-d$.
Throughout our analysis we will closely adhere to the approach outlined in
Ref.~\cite{bzj}. Additional details can also be found in Ref.~\cite{qft},
Ch.~36.

We consider a system with a hypercubic geometry, with linear dimension~$L$ and
periodic boundary conditions. It is represented by the following
Landau--Ginzburg--Wilson Hamiltonian in momentum space,
\begin{equation}
 {\cal H}(\phi_{\bf k})/k_B T =
 \frac{1}{2} \sum_{\bf k} \sum_i (k^\sigma + r_0) 
             \phi_{i,{\bf k}}\phi_{i,-{\bf k}} +
 \frac{1}{4!}\frac{1}{L^d}u_0 
   \sum_{{\bf k}_1,{\bf k}_2,{\bf k}_3} \sum_{i,j}
   \phi_{i,{\bf k}_1}\phi_{i,{\bf k}_2}\phi_{j,{\bf k}_3}
   \phi_{j,-{\bf k}_1-{\bf k}_2-{\bf k}_3} \;,
\label{eq:hamil}
\end{equation}
where the factor $k^\sigma$ ($0<\sigma<2$) arises from the isotropic long-range
interactions. Compared to the SR case, it leads to the general replacement $k^2
\to k^\sigma$ in all propagators.  The indices $i,j$ refer to the components of
the field. It has been well established that this system belongs to the same
universality class as a discrete spin model with algebraically decaying
interactions; in particular the $\varepsilon$-expansion
results~\cite{fmn,suzuki72} for the critical exponents have been confirmed (see
Refs.~\cite{fss_ud,iuu} and references therein).
Due to the finite geometry all components of the wave vectors are integer
multiples of $\frac{2\pi}{L}$.  The sums run to infinity, which corresponds to
a vanishing lattice spacing~$a$; however, the ratio $L/\xi$ is finite, whereas
$L/a$ and $\xi/a$ are both sent to infinity~\cite{bzj}.  Expectation values are
computed from a partition function in which (\ref{eq:hamil}) is replaced by an
effective Hamiltonian, consisting of the exactly calculated ${\bf k}={\bf 0}$
(homogeneous) mode contribution and a perturbatively calculated part, which
contains the contribution of all nonzero modes.  Only the latter contribution
is affected by the changeover to long-range interactions, cf.\ 
Ref.~\cite{fss_ud}. To one-loop order it consists of a shift of the critical
temperature and a renormalization of the coupling constant. Higher operators do
not contribute at this order.  We introduce the dimensionless coupling constant
$g_0=\mu^{-\varepsilon}u_0$ and work in the system of units where $\mu=1$. The
parameter $r_0$ is split into $r_{0c} + t$, where $t = r_0-r_{0c} \propto
T-T_c$ and we require $t \geq 0$.  The RG calculations are carried out in the
minimal subtraction scheme~\cite{bgz,amit}, where it is a crucial ingredient of
the calculation that, despite the quantization of all momenta, the UV
divergences are taken care of by the bulk renormalization constants. As we do
not go beyond one loop, we have ignored the field renormalization constant.

The leading contribution to the shift of $T_c$ is
\begin{equation}
 \frac{n+2}{6}g_0\frac{1}{L^d}\sum_{{\bf k}}{\vphantom{\sum}}^\prime 
                              \frac{1}{|{\bf k}|^\sigma + t} \;,
\label{eq:tc-shift}
\end{equation}
where the prime indicates that the ${\bf k}={\bf 0}$ mode is omitted from the
sum. In the Schwinger parametrization this can be rewritten as
\begin{equation}
 L^{-d}\int_0^\infty ds\, e^{-st}
 \left[ \sum_{m_1=-\infty}^{\infty}\cdots\sum_{m_d=-\infty}^{\infty}
        e^{-s(m^2)^{\sigma/2}(2\pi/L)^{\sigma}} -1 \right] \;,
\label{eq:div-int}
\end{equation}
where $m^2=\sum_{i=1}^{d} m_i^2$ and we have omitted the prefactor
$\frac{n+2}{6}g_0$.  For $d \geq \sigma$, the UV~divergence in
Eq.~(\ref{eq:tc-shift}) is reflected by the divergence of the integral at small
$s$. Thus we isolate this divergence by rewriting Eq.~(\ref{eq:div-int}) as
\begin{equation}
 \frac{L^{\sigma-d}}{(2\pi)^\sigma} I_1(d,\sigma,t)
 + \frac{L^{\sigma-d}}{(2\pi)^\sigma} S_{d-1}\frac{1}{\sigma}
   \Gamma(\frac{d}{\sigma})
   \int_0^\infty du\, u^{-d/\sigma} e^{-ut(L/2\pi)^\sigma} \;,
\label{eq:div-isol}
\end{equation}
with
\begin{equation}
 I_1(d,\sigma,t) \equiv 
 \int_0^\infty du\, e^{-ut(L/2\pi)^\sigma}
 \left[ \sum_{m_1}\cdots\sum_{m_d}
        e^{-u(m^2)^{\sigma/2}} -1 - S_{d-1} \frac{1}{\sigma}
 \Gamma(\frac{d}{\sigma})u^{-d/\sigma} \right] \;,
\end{equation}
which is finite. $S_{d-1}=2\pi^{d/2}/\Gamma(\frac{d}{2})$ is the surface area
of a $d$-dimensional unit sphere.  The second term in Eq.~(\ref{eq:div-isol})
can be continued analytically for $\Re(d) \geq \sigma$ and has a simple pole
for $d=2\sigma$ ($\varepsilon=0$). Upon expansion around this pole we find for
Eq.~(\ref{eq:div-isol})
\begin{equation}
 - \frac{2t}{(4\pi)^\sigma\Gamma(\sigma)\varepsilon}
 + \frac{t}{(4\pi)^\sigma\Gamma(\sigma)}
   \left\{ \frac{2}{\sigma}\ln t - \left[\ln 4\pi + \Psi(\sigma)\right]\right\}
 + \frac{1}{(2\pi L)^\sigma} I_1(2\sigma,\sigma,t) + {\cal O}(\varepsilon)\;,
\end{equation}
where $\Psi(\sigma)$ denotes the Digamma function.  The addition of the
$\phi^2$-insertion counterterm leads to the replacement of $t$ by
$tZ_{\phi^2}$. To one-loop order, the renormalization constant has the same
form as for SR interactions~\cite{yamsuz}; the pole is canceled and the shifted
reduced temperature is given by
\begin{equation}
 \tilde{t} =
 t + \frac{n+2}{6\sigma}\hat{g}_0 t \ln t 
   + \frac{2^\sigma}{12}(n+2)\Gamma(\sigma)\hat{g}_0 \frac{1}{L^{\sigma}} 
     I_1(2\sigma,\sigma,t)
   + {\cal O}(\hat{g}_0^2) \;,
\label{eq:t-renorm}
\end{equation}
with 
$\hat{g}_0 = 2[(4\pi)^\sigma \Gamma(\sigma)]^{-1} g_0 \{ 1 + \frac{1}{2}
\varepsilon [\ln 4\pi + \Psi(\sigma)] + {\cal O}(\varepsilon^2) \}$.
The renormalized coupling constant~$g$ is calculated in a similar fashion. The
leading finite-size contribution is given by $-\frac{n+8}{6}g_0^2 L^{-d}
\sum_{{\bf k}}{\vphantom{\sum}}^\prime (|{\bf k}|^\sigma + t)^{-2}$,
which has a UV divergence for $\Re(d) \geq 2\sigma$. The $1/\varepsilon$ pole
is canceled by the counterterm, where the renormalization constant to one-loop
order is given by $Z_g = 1 + \frac{n+8}{6\varepsilon}\hat{g}$. After some
algebra, we find for the renormalized coupling constant
\begin{equation}
   g = 
   g_0 \left[ 1 + \frac{n+8}{6\sigma}\hat{g}_0(1+\ln t) 
              - \frac{n+8}{12}\hat{g}_0 \frac{\Gamma(\sigma)}{\pi^\sigma} 
              I_2(2\sigma,\sigma,t) + {\cal O}(\hat{g}_0^2) \right] \;,
\label{eq:g-renorm}
\end{equation}
with
\begin{equation}
 I_2(d,\sigma,t) \equiv 
 \int_0^\infty du\, ue^{-ut(L/2\pi)^\sigma}
 \left[ \sum_{m_1}\cdots\sum_{m_d}
        e^{-u(m^2)^{\sigma/2}} -1 - S_{d-1} \frac{1}{\sigma}
 \Gamma(\frac{d}{\sigma})u^{-d/\sigma} \right] \;.
\end{equation}
Equations (\ref{eq:t-renorm}) and~(\ref{eq:g-renorm}) suffice to calculate the
finite-size scaling functions close to criticality to ${\cal O}(\varepsilon)$.
To this order, the fixed-point value of $\hat{g}_0$ only differs from the SR
case in the definition of $\varepsilon$, $\hat{g}_0^* =
\frac{6\varepsilon}{n+8} + {\cal O}(\varepsilon^2)$. We are now able to
calculate $\tilde{t}L^{d/2}g^{-1/2}$ at the fixed point, which, as we shall see
shortly, is the parameter appearing in the finite-size scaling functions.  This
also provides a simple consistency check for our calculations, since all
factors $\ln L$ have to disappear upon introduction of the appropriate scaling
variable $y \equiv tL^{1/\nu}$.  We find $1/\nu = \sigma -
\frac{n+2}{n+8}\varepsilon + {\cal O}(\varepsilon^2)$, which indeed agrees with
Ref.~\cite{fmn}, and the final expression is
\begin{eqnarray}
   \left. \tilde{t}L^{d/2}g^{-1/2} \right|_{{\rm f.p.}} =
   \frac{1}{\sqrt{g^*_0}} & &
   \left[ y - \frac{1}{2\sigma} \varepsilon y 
            + \frac{n-4}{2\sigma(n+8)} \varepsilon y \ln y
            + \frac{1}{4}\varepsilon y \frac{\Gamma(\sigma)}{\pi^\sigma} 
              I_2(2\sigma,\sigma,yL^{-1/\nu}) \right.  \nonumber \\
  & &
  \left. \phantom{[} 
            + \frac{2^{\sigma-1}(n+2)}{n+8}\varepsilon \Gamma(\sigma) 
              I_1(2\sigma,\sigma,yL^{-1/\nu})
            + {\cal O}(\varepsilon^2) \right] \;.
\end{eqnarray}
We are particularly interested in the amplitude ratio $Q$ at criticality, for
$n=1$. The even moments of the magnetization distribution are calculated from
$\langle (\phi^2)^p \rangle = \int_{-\infty}^{+\infty} d\phi\, (\phi^2)^p
\exp[-S(\phi)] / \int_{-\infty}^{+\infty} d\phi\, \exp[-S(\phi)]$,
with $S(\phi)=L^d(\frac{1}{2}\tilde{t}\phi^2+\frac{1}{4!}g\phi^4)$.
We carry out the rescaling $\phi \to (L^d g)^{-1/4}\phi$ and expand in terms
of the parameter $x \equiv \tilde{t} L^{d/2} g^{-1/2}$. Elementary algebra
leads to
\begin{equation}
 Q = \frac{4\Gamma^2(\frac{3}{4})}{\Gamma^2(\frac{1}{4})}
     \left[ 1 + \left( 4\frac{\Gamma(\frac{3}{4})}{\Gamma(\frac{1}{4})}
                       - \frac{1}{2}
                         \frac{\Gamma(\frac{1}{4})}{\Gamma(\frac{3}{4})}\right)
                       \sqrt{6}x
              + \left( 13\frac{\Gamma^2(\frac{3}{4})}{\Gamma^2(\frac{1}{4})}
                       + \frac{1}{16}
                         \frac{\Gamma^2(\frac{1}{4})}{\Gamma^2(\frac{3}{4})}
                       - 2\right) 6x^2
              + {\cal O}(x^3) \right] \;.
\label{eq:q-expan}
\end{equation}
At criticality, $y=0$ and $x$ takes the value
\begin{equation}
 x_0 = \sqrt{\varepsilon}
       \left\{\frac{1}{2}\frac{n+2}{\sqrt{3(n+8)}} 
             \sqrt{\frac{\Gamma(\sigma)}{\pi^\sigma}}
             I_1(2\sigma,\sigma,0) + {\cal O}(\varepsilon)
       \right\} \;.
\label{eq:x0}
\end{equation}
A comparison to Eq.~(3.33) in Ref.~\cite{bzj} shows that $x_0$ only differs
from the SR case by a redefinition of $\varepsilon$, a geometric factor, and
the integral $I_1$ and thus we see that the singular structure is preserved in
the generalization to long-range forces. For completeness we remark that one
may also calculate $x_0$ by carrying out all manipulations at criticality. This
permits a different parametrization and leads to the same expression for $x_0$,
in which $I_1(2\sigma,\sigma,0)$ is replaced by
$\hat{I}_1(2\sigma,\sigma)/\Gamma(\sigma/2)$, where $\hat{I}_1(d,\sigma)
\equiv\int_0^\infty du\, u^{\sigma/2-1} [ ( \sum_{m=-\infty}^{\infty} e^{-um^2}
)^d - 1 - (\pi/u)^{d/2} ]$.
As a side note, we conjecture $\hat{I}_1(4,2)$, which has been evaluated
numerically in Refs.~\cite{bzj,qft,dohm}, to be exactly equal to $-8\ln 2$. For
lower dimensionalities, the upper critical dimension shifts toward smaller
values of $\sigma$, and numerical evaluation yields: $I_1(d,d/2,0)=-2.92,
-3.900, -4.8227$ for $d=1,2,3$, respectively. Substitution into Eqs.\ 
(\ref{eq:x0}), (\ref{eq:q-expan}) suggests that for each of these values a
reasonable convergence may be expected for $\varepsilon < 1$.

\begin{figure}
\leavevmode
\centering
\epsfxsize 10cm
\epsfbox{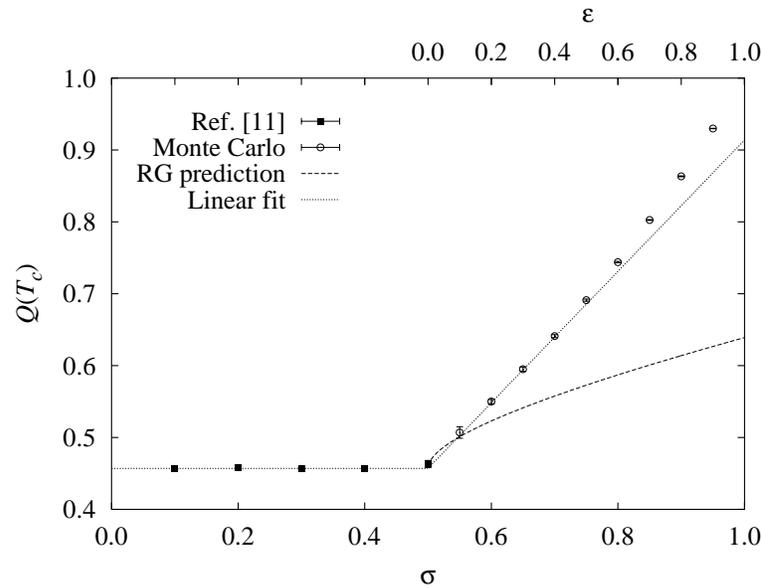}
\caption{The critical value of the amplitude ratio $Q$ as a function of the
decay parameter $\sigma$ for $d=1$. The corresponding values of $\varepsilon$
are shown at the upper axis. The data in the regime $0 < \sigma \leq 0.5$ are
taken from Ref.~\protect\cite{fss_ud}.}
\label{fig:q-1d}
\end{figure}

\begin{figure}
\leavevmode
\centering
\epsfxsize 10cm
\epsfbox{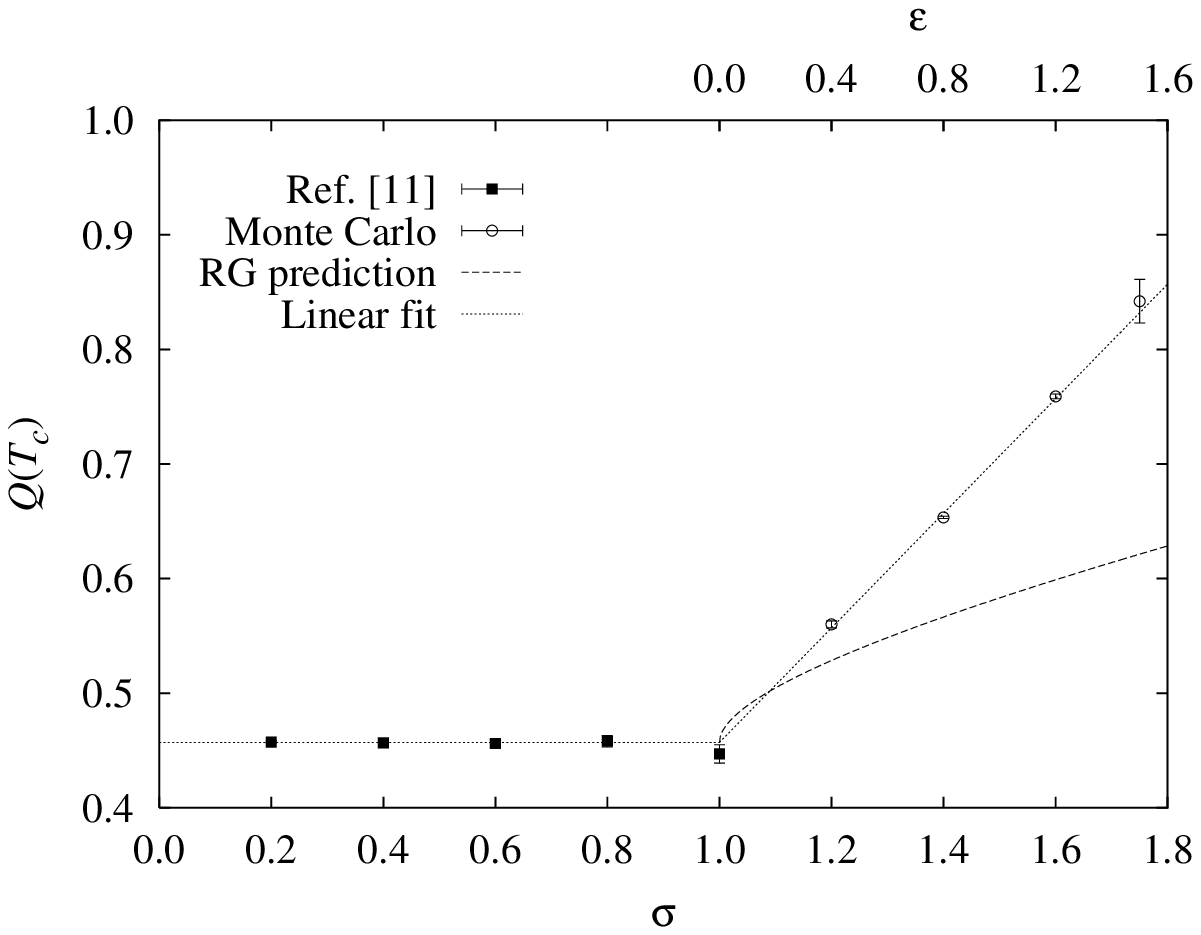}
\caption{The analog of Fig.~\protect\ref{fig:q-1d} for $d=2$.}
\label{fig:q-2d}
\end{figure}

In order to verify these predictions, we have carried out extensive Monte Carlo
simulations of spin models with $n=1$ and algebraically decaying interactions,
for $d=1,2$. Accurate results could be obtained by means of an efficient
cluster algorithm~\cite{ijmpc}.  We investigated system sizes $10 \leq L \leq
150\,000$ for $d=1$ and $4 \leq L \leq 400$ for $d=2$, for several values of
the decay parameter $\sigma$. These were chosen such that $0 < \varepsilon
\lesssim 1$, where it should be noted that for very small $\varepsilon$ the
analysis is hampered by strong corrections to scaling.  Simulational details
can be found in Ref.~\cite{fss_ud}, where the classical regime $0 < \sigma \leq
d/2$ is discussed.  The numerical results were analyzed using an expression
similar to Eq.~(13) in Ref.~\cite{fss_ud}, $Q_L(T) = Q + r_1 t L^{y_t} + r_2
t^2 L^{2y_t} + \cdots + s_1 L^{y_i} + \cdots$.
Here $y_t$ and $y_i$ are the thermal and leading irrelevant exponent,
respectively, $r_i$ and $s_i$ are nonuniversal coefficients and the ellipses
denote higher-order terms. An extensive analysis of the data will be presented
elsewhere.  The resulting estimates for $Q(T_c)$ are shown in
Figs.~\ref{fig:q-1d},~\ref{fig:q-2d}. For the one-dimensional case
(Fig.~\ref{fig:q-1d}), the region $0.1 \leq \varepsilon \leq 0.9$ has been
covered. For $\sigma \to 1$, the data points approach $Q(T_c)=1$, in agreement
with the occurrence of a Kosterlitz--Thouless transition at
$\sigma=1$~\cite{kosterlitz76}. However, for $0 < \varepsilon \lesssim 0.5$ the
numerical results are described by a perfectly linear dependence on
$\varepsilon$, in strong contrast with the predicted square-root behavior. This
discrepancy is reinforced by the two-dimensional results (Fig.~\ref{fig:q-2d}),
which are also well described by a linear relation for $0 < \varepsilon
\lesssim 1.2$. For larger $\sigma$, the error bars increase, signaling the
crossover to short-range criticality. In contrast, it should be noted that for
the critical exponents good agreement with the RG predictions has been reached
for at least $0 < \varepsilon \lesssim 0.5$, both for $d=1$ and
$d=2$~\cite{iuu}.

In summary, we have calculated universal finite-size scaling functions to
second order in $\sqrt{\varepsilon}$ for systems with algebraically decaying
interactions. These calculations are essentially a generalization of those for
systems with short-range interactions~\cite{bzj} and exhibit the same singular
dependence on $\varepsilon$. Subsequently, we have compared our results to
accurate simulations for one- and two dimensional systems belonging to the same
universality class as the field-theoretical Hamiltonian. The presence of
long-range interactions offers the advantage that $\varepsilon$ is a continuous
parameter, so that one can reach the regime where the convergence of the
$\varepsilon$-expansion has not to be doubted. Nevertheless, no agreement is
found: the numerical data exhibit a linear rather than a square-root dependence
on $\varepsilon$. Currently, we do not have an explanation for this striking
discrepancy. Although higher-order terms might yield some improvement, it is
difficult to envisage that this would fully resolve the problems. It would be
very remarkable if the apparent linear variation over such a wide range in
Figs.\ \ref{fig:q-1d} and~\ref{fig:q-2d}, which includes the point where our
$\sqrt{\varepsilon}$-expansion starts from, were accidental.  In
Ref.~\cite{hilfer}, it has been suggested that $Q$ contains {\em
nonuniversal\/} contributions, depending on the cutoff used in the integration
of the order parameter probability distribution. However, apart from the
validity of this suggestion, it is difficult to envisage how this would lead to
the (dis)appearance of a square-root contribution in the
$\varepsilon$-expansion.  Furthermore, it is an open question to what extent
the breakdown of the field-theoretic description of finite-size scaling for $d
\geq 4$~\cite{dohm} influences the nature of the $\varepsilon$-expansion. We
feel that an understanding of these problems is of some significance for the
understanding of finite-size scaling of critical phenomena in general.

\acknowledgements
It is a pleasure to thank Prof.\ K. Binder and Prof.\ H. Bl\"ote for useful
comments.

\vspace{3mm}
\paragraph*{Note:}
After completion of this work, we became aware of Ref.~\cite{korutcheva91},
which contains a different generalization of the results of Ref.~\cite{bzj}. It
allows for the presence of both short-range and long-range interactions, but
imposes the additional restriction that $\sigma = 2 - {\cal O}(\varepsilon)$.

\begin{references}
\bibitem{barber} M.~N. Barber, in {\it Phase Transitions and Critical
  Phenomena}, Vol.~8, edited by C. Domb and J.~L. Lebowitz (Academic, London,
  1983); V. Privman (ed.), {\it Finite Size Scaling and Numerical Simulation of
  Statistical Systems\/} (World Scientific, Singapore, 1990). 
\bibitem{fisher-fss} M.~E. Fisher, in {\it Critical Phenomena}, Proc. 51st
  Enrico Fermi Summer School, Varenna, Italy, edited by M.~S. Green (Academic,
  N.Y., 1971).
\bibitem{brezin} E. Br\'ezin, J. Phys. (Paris) {\bf 43}, 15 (1982).
\bibitem{privman84} V. Privman and M.~E. Fisher, Phys. Rev. B {\bf 30}, 322
  (1984).
\bibitem{sp85} S. Singh and R.~K. Pathria, Phys. Rev. B {\bf 31}, 4483 (1985).
\bibitem{bzj} E. Br\'ezin and J. Zinn-Justin, Nucl. Phys. {\bf B257} [FS14],
  867 (1985).
\bibitem{rudnick} J. Rudnick, H. Guo, and D. Jasnow, J. Stat. Phys. {\bf 41},
  353 (1985).
\bibitem{esser} A. Esser, V. Dohm, M. Hermes, and J.~S. Wang, Z. Phys. B {\bf
  97}, 205 (1995).
\bibitem{binder-cum} K. Binder, Z. Phys. B {\bf 43}, 119 (1981).
\bibitem{fmn} M.~E. Fisher, S.-k. Ma, and B.~G. Nickel, Phys. Rev. Lett. {\bf
  29}, 917 (1972).
\bibitem{suzuki72} M. Suzuki, Y. Yamazaki, and G. Igarashi, Phys. Lett. {\bf
  42A}, 313 (1972).
\bibitem{lr-spherical} M.~E. Fisher and V. Privman, Comm. Math. Phys. {\bf
  103}, 527 (1986);
  J.~G. Brankov and N.~S. Tonchev, J. Stat. Phys. {\bf 52}, 143 (1988);
  S. Singh and R.~K. Pathria, Phys. Rev. B {\bf 40}, 9238 (1989).
\bibitem{qft} J. Zinn-Justin, {\it Quantum Field Theory and Critical
  Phenomena}, third edition (Clarendon, Oxford, 1996).
\bibitem{fss_ud} E. Luijten and H.~W.~J. Bl\"ote, Phys. Rev. Lett. {\bf 76},
  1557, 3662(E) (1996); Phys. Rev. B {\bf 56}, 8945 (1997).
\bibitem{iuu} E. Luijten, {\it Interaction Range, Universality and the
  Upper Critical Dimension\/} (Delft University Press, Delft, 1997).
\bibitem{bgz} E. Br\'ezin, J.~C. Le Guillou, and J. Zinn-Justin, in {\it Phase
  Transitions and Critical Phenomena}, Vol.~6, edited by C. Domb and
  M.~S. Green (Academic, London, 1976).
\bibitem{amit} D.~J. Amit, {\it Field Theory, The Renormalization Group, and
  Critical Phenomena}, second edition (World Scientific, Singapore, 1984).
\bibitem{yamsuz} Y. Yamazaki and M. Suzuki, Prog. Theor. Phys. {\bf 57}, 1886
  (1976).
\bibitem{dohm} X.~S. Chen and V. Dohm, Int. J. Mod. Phys. C {\bf 9}, 1007
  (1998).
\bibitem{ijmpc} E. Luijten and H.~W.~J. Bl\"ote, Int. J. Mod. Phys. C {\bf 6},
  359 (1995).
\bibitem{kosterlitz76} J.~M. Kosterlitz, Phys. Rev. Lett. {\bf 37}, 1577
  (1976).
\bibitem{hilfer} R. Hilfer, Z. Phys. B {\bf 96}, 63 (1994); R. Hilfer and
  N.~B. Wilding, J. Phys. A {\bf 28}, L281 (1994).
\bibitem{korutcheva91} E.~R. Korutcheva and N.~S. Tonchev, J. Stat. Phys. {\bf
  62}, 553 (1991).
\end{references}
\end{document}